# Design and performance of a TDC ASIC for the upgrade of the ATLAS Monitored Drift Tube detector

Yu Liang[a,b], Jinhong Wang[b], Xiong Xiao[b], Alessandra Pipino[c], Yuxiang Guo[a, b],

Qi An[a], Andrea Baschirotto[c], J. W. Chapman[b], Tiesheng Dai[b], Marcello de Matteis[c], Markus Fras[d], Oliver Kortner[d], Hubert Kroha[d], Federica Resta[c], Robert Richter[d], Lei Zhao[a], Zhengguo Zhao[a], Bing Zhou[b], Junjie Zhu[b]

[a]*State Key Laboratory of Particle Detection and Electronics, University of Science and Technology of China, Hefei 230026, China*
[b]*Department of Physics, University of Michigan, Ann Arbor, MI, 48109, USA*
[c]*Department of Physics, University of Milan-Bicocca, 20126 Milano, Italy*
[d]*Max-Planck-Institut for Physics, 80805 Munich, Germany*

ABSTRACT

We present the prototype of a time-to-digital (TDC) ASIC for the upgrade of the ATLAS Monitored Drift Tube (MDT) detector for high-luminosity LHC operation. This ASIC is based on a previously submitted demonstrator ASIC designed for timing performance evaluation, and includes all features necessary for the various operation modes, as well as the migration to the TSMC 130 nm CMOS technology. We present the TDC design with the emphasis on added features and performance optimization. Tests of the timing performance demonstrate that this ASIC meets the design specifications. The TDC has a bin size of about 780 ps, and a timing bin variations within ±40 ps for all 24 channels with leading and trailing edge digitization, while the power consumption has been limited to 250 mW, corresponding to a consumption of about 5.2 mW per edge measurement.

## 1. Introduction

The ATLAS muon spectrometer covers the outer part of the ATLAS experiment at the LHC collider at CERN. The measurement of the muon track coordinates relies on three stations of Monitored Drift Tubes (MDTs) chambers and is designed to determine the transverse momentum of the track with an accuracy of 10% at 1 TeV/c [1]. Each chamber is equipped with up to 18 electronics cards called "mezzanine cards" containing three custom-designed monolithic Amplifier/Shaper/Discriminator (ASD) chips [2] and one Time-to-Digital Converter (TDC) chip [3], where each card can read out 24 tubes. All mezzanine cards on a chamber are controlled and read out by an on-chamber data-acquisition (DAQ) element referred to as a Chamber Service Module (CSM). The CSM multiplexes data from the corresponding mezzanine cards and sends these data via an optical fiber to a rear-end processor, called the MDT Readout Driver (MROD) [4-5].

The current readout system of the MDT chambers was designed to cope with the original LHC design luminosity of $10^{34}$ $cm^{-2}s^{-1}$. Starting in 2025, the luminosity of the high-luminosity LHC (HL-LHC) will be increased by a factor of 5-7, which requires a substantial upgrade of the trigger and readout electronics of the MDT detector, as well as of all other ATLAS sub-detectors. The targeted ATLAS trigger accept rate and latency at the first trigger level (L1) for future HL-LHC runs are 1 MHz and 10 μs, respectively. The present L1 muon trigger is exclusively based on dedicated trigger chambers with high time resolution but modest spatial resolution (about 2 cm), limiting the momentum resolution of the trigger candidate tracks and thus allowing a large fraction of muons with momenta substantially below the threshold to pass the L1 trigger condition. For the HL-LHC upgrade, the MDT information will be available at L1. A large fraction of these unwanted L1 triggers could be suppressed using the high accuracy of the MDT coordinates (about 100 μm) for a better determination of the muon momentum. The design of the TDC must minimize the latency for the incoming data due to the increasing demand for intermediate storage, processing and formatting with increasing latency.

Figure 1 shows the MDT trigger and readout chains for HL-LHC runs. The MDT on-chamber electronics will send all muon hits to the circuitry in the ATLAS counting room. In a first step, signals from MDT tubes are processed by an ASD chip [6-8] and subsequently time-digitized by the TDC chip. These time measurements are transmitted through the CSM to a MDT Data Processor Board, where relevant hits are extracted out of the raw data stream. For this hit extraction, timing information from the fast trigger chambers (RPCs in the barrel and TGCs in the endcap) is used to define a time $t_0$ when the triggering muon actually passed the chamber. The difference between the arrival time of the first signal at the central wire and $t_0$ is equal to the hit drift time in the tube and can be converted into the distance of the track from the wire. These "precision coordinates" of all matched hits are then processed. Segment finder and track fitter algorithms are applied to determine the muon momentum. Time information for hits matching the L1 trigger time window (L1 accept) is stored for transmission to the FELIX readout system [9] after receiving the first-level trigger accept signal.

This paper describes the design of a fully-featured TDC for the MDT electronics upgrade, as well as performance characterization of a recent prototype using the TSMC 130 nm CMOS process. The paper is organized as follows: Section 2 introduces the overall TDC design with an emphasis on newly added features and power optimization, Section 3 presents the results from performance evaluation of a recent prototype and Section 4 gives conclusions.

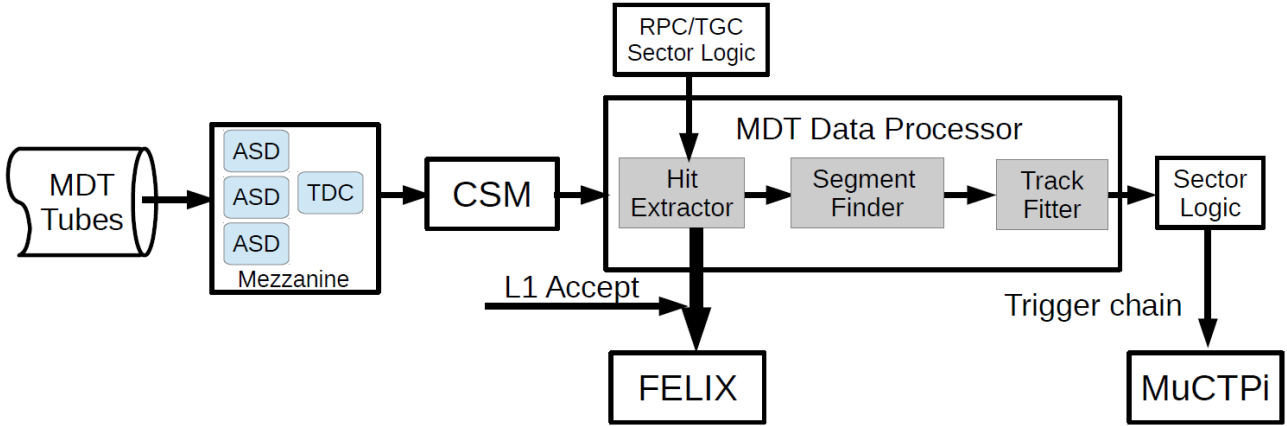

Fig. 1: Block diagram of the MDT trigger and readout electronics for HL-LHC runs. All muon hits will be sent by the on-chamber electronics (ASD, TDC and CSM) to the off-chamber electronics in the ATLAS counting room. Relevant hits will be extracted based on the time information provided by RPC and TGC trigger chambers. Segment finder and track fitter algorithms are applied on matched hits to determine the accurate momentum of the incoming muon. The fitted muon track information will be sent to the ATLAS Sector Logic and muon Central Track (trigger) Processor (MuCTPi). All hits matching the L1 trigger time window will be sent to the FELIX readout system.

## 2. Design of the TDC prototype

The TDC is responsible for the time digitization of both leading and trailing edges of discriminated signals from the ASD. This time digitization is the basis for all subsequent trigger and readout processing. To maintain a position resolution of about 100 µm per tube, it is required that the contribution of front-end electronics to the tube position resolution be less than 20 µm [4]. Since the average drift velocity is 20.7 µm/ns, the TDC will use a least count of 0.78 ns and a digitization of 17 bits to cover the time of one LHC orbit cycle (~102.4 us).

The TDC has two modes of data recording: (a) the "edge mode", where times for leading and trailing edges are recorded separately and (b) the "pair mode" in which the arrival time of the leading edge plus time difference between both edges are recorded.

Furthermore, there are two operational modes of the TDC depending on whether or not a trigger accept signal is required to select hits for output, denoted as "triggerless" or "triggered" modes, respectively. In the triggerless mode, no trigger request is needed and the time measurements are shifted out of the TDC immediately, as long as the bandwidth of the link to the CSM can copes with the respective data rate. In the triggered mode, time measurements are buffered waiting for the trigger accept signal which is used in a time-matching algorithm to select hits of interest for outputs. Due to the expected L1 trigger rate of at least 1 MHz for future HL-LHC runs, the triggerless mode will be the default operation mode. The triggered mode, on the other hand, maintains full compatibility with the current MDT TDC chip (the AMT ASIC [3]) and is also useful for chamber testing and test beam studies. Both options are implemented in this TDC design.

The overall power consumption is another critical design parameter. Our target is a consumption below 360 mW, which is the power consumption of the current AMT ASIC [4].

A TDC demonstrator prototype using the GlobalFoundries 130 nm CMOS process has previously been designed and fabricated in 2016 [10]. The main purpose of this prototype was to develop and validate the overall TDC architecture. Studies of the prototype indicate that the design meets the time measurement requirement. However, the demonstrator prototype does not contain all features for its use at HL-LHC runs. The total power consumption was found to be 310 mW even though only the triggerless mode was implemented. In addition, we migrated the design from the Global Foundries 130 nm CMOS process to the 130 nm process from TSMC.

In this paper, we present the design of a new TDC prototype containing all HL-LHC related design features. Additional work on the TDC included the technology migration to the TSMC 130 nm CMOS process and circuit optimization in the new technology. The overall TDC schematic diagram is shown in Figure 2. There are 24 inputs and two serial outputs, each with a maximum line rate of 320 Mbps [11], and a reference clock provided by the 40 MHz LHC bunch crossing clock. The 24 input channels are arranged symmetrically on the two sides, with 12 channels each. We divide the design into four main building blocks:

(a) The extended Phase-Lock Loop unit (ePLL) that provides two copies of 320 MHz and one 160 MHz clock with programmable phases;
(b) The time digitization unit that digitizes the arrival times of leading and trailing edges;
(c) The TDC logic unit that processes the digitized time from all 24 channels, and transmits the formatted data to the CSM with configurable output rates;
(d) The configuration and monitoring unit is responsible for TDC configuration, communication with the ASD ASICs and monitoring relevant chip parameters.

In the following sections, we will focus on the details of the first three items as the fourth item represents a conventional logic implementation of the JTAG interface towards the ASD ASICs as well as to CSM.

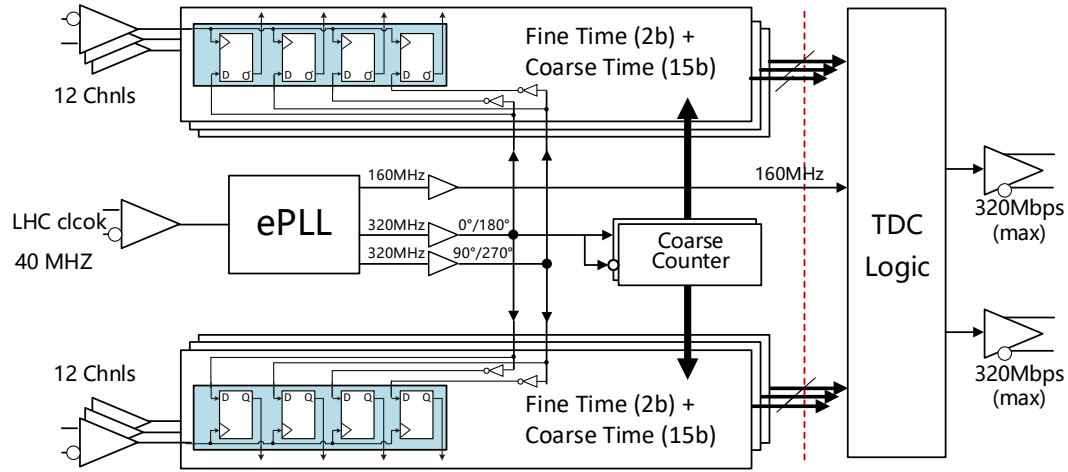

Fig. 2: The schematic diagram of the TDC ASIC. Both leading and trailing edges of the discriminated detector signals are digitized using the clocks provided by the ePLL unit. Digitized time data from all 24 channels are then sent to the TDC logic unit where they are formatted and sent to CSM via two output lines at a maximum date rate of 320 Mbps per line.

*2.1 The ePLL unit*

The design of the ePLL was originated from the CERN microelectronics group [12], in which the input reference clock is selectable among 40/80/160 MHz and it generates three copies of 320 MHz and 160 MHz clocks for outputs, with a phase programmable step of 22.5° and 11.25°, respectively. The design was ported to the TSMC 130 nm CMOS technology utilized by the TDC design [13] and a few modifications were also made, in which only one copy of 160 MHz and two copies of 320 MHz are reserved. Due to the lower supply voltage of 1.2 V instead of 1.5 V and the simplifications made in the ePLL functionalities, the total power consumption has been reduced to 12.5 mW compared to 31 mW from the previous prototype design.

*2.2. Time digitization unit*

The TDC utilizes a combination of coarse time and fine time for the time measurement, where the dynamic range is covered by the coarse-time with a LSB of 3.125 ns. The fine time is obtained from a subdivision of the coarse time by a factor of 4, leading to a LSB of 0.78 ns, which is the target value for the drift time measurement in the MDT tubes. The coarse time measurement is based on two coarse counters running at inverting phases of 320 MHz, while the fine time measurement is achieved by phase interpolation of four phases of 320 MHz clocks, with a difference of 90° relative to each other, i.e. the phases are 0°/90°/180°/270°. We make use of both rising and falling edges of two 320 MHz clocks, and therefore only a phase difference of 90° between the two clocks is required. It is mandatory to keep a 50% duty cycle of each clock to optimize linearity.

In the implementation of the time digitization unit, we follow the architecture of the demonstrator prototype and use the input hit signal to sample clocks instead of sampling the input hit signal with multiple phases of clocks. Such choice avoids crossings between multiple clock domains and also minimizes the number of sampling registers for implementation. A detailed description of the TDC time unit can be found in [10].

The sampling register is a crucial component of the time digitization unit. In the prior work we improved the linearity of the TDC time measurements by reducing the uncertainty of their setup and hold times. Strategies for mitigating ambiguous bins were also implemented. As a result, the inequality among the 4 fine time steps inside a coarse time bin of 3.125 ns is kept within ±40 ps, i.e. 5% of the TDC bin size. This design method is also implemented in the present TDC prototype.

We further reduced the power dissipation of the TDC by replacing the True single Clocked Latches (TSPC) used in the demonstrator prototype with a combination of both dynamic and static latches, as shown in Figure 3(a). TSPC offers fast sampling, whereas the bias of internal nodes (such as node c) would drift significantly over the time period of two consecutive hits, which can be of the order of several milliseconds. This drift may put the output transistors into the conduction region, increasing the total power consumption. Power consumption can be reduced by replacing the dynamic output latch with a static latch, denoted as Pulse-Triggered Register (PTR) as shown in Figure 3(b). A flag signal (the complementary pair of Ready and Ready_inv) is introduced so that once the fine time measurement is captured, this signal will turn the output stage into a locking state until the next hit arrives at the input. Simulation shows that the average power per register is reduced from 50 µW to 125 nW under the typical technology corner. There are a total of 192 sampling registers in the TDC, corresponding to a power saving of about 19 mW.

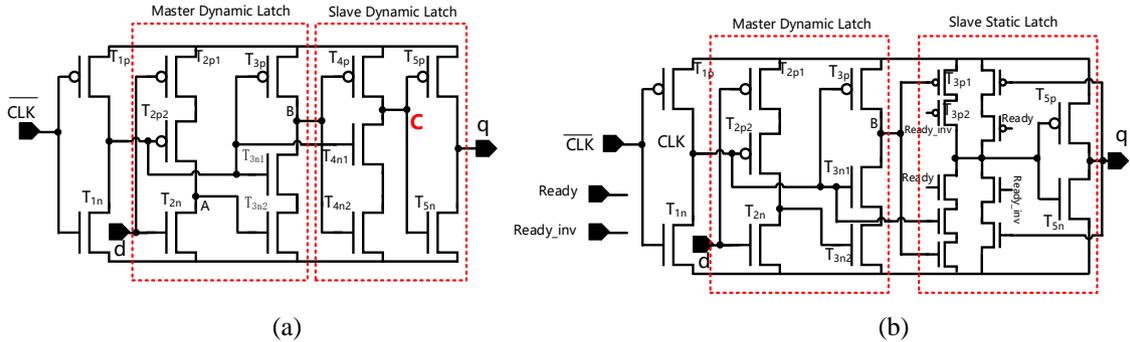

Fig 3: Schematic diagrams of the sampling register: (a) the True single Clocked Latches (TSPC) used for the demonstrator prototype and (b) the Pulse-Triggered Register (PTR) of the new prototype using a slave static latch to reduce power consumption.

*2.3. The TDC Logic*

The TDC logic processes the time measurements from each TDC channel, and assembles them in the correct format depending on the operation modes. A block diagram is shown in Figure 4. There are 24 copies of the channel logic, corresponding to the data processing inside 24 TDC channels plus shared functions such as Trigger-Time-and Control (TTC), monitoring and configuration. The signal flow in triggerless and triggered operation is illustrated by solid and dashed lines, respectively. The two operation modes share common blocks such as the Hit builder, Channel FIFO and readout FIFO, while each mode has its own unique block for specific operations.

The format of the TDC's raw time measurements is summarized in Table 1, in which the 24-bit raw word for the edge mode comprises 5 bits for the channel identification number (ID), 2 bits for the operation mode (with leading edge mode denoted as "00", trailing edge mode denoted as "01", and pair mode denoted as "11") and 17 bits for each edge time measurement. In pair mode, 8 bits are used for the pulse width, and the resolution of the bin width measurement can be adjusted by truncating the 8 bits from the full 17-bit measurement. The pair mode leads to a substantial reduction of the data volume compared to sending out the full time information for two edges, as the full dynamic range of 17 bit is only needed for the leading edge, while the time difference to the trailing edge can be expressed with only 8 bits. The raw packets are arranged in multiple of 8 bits, making it natural for 8b/10b encoding.

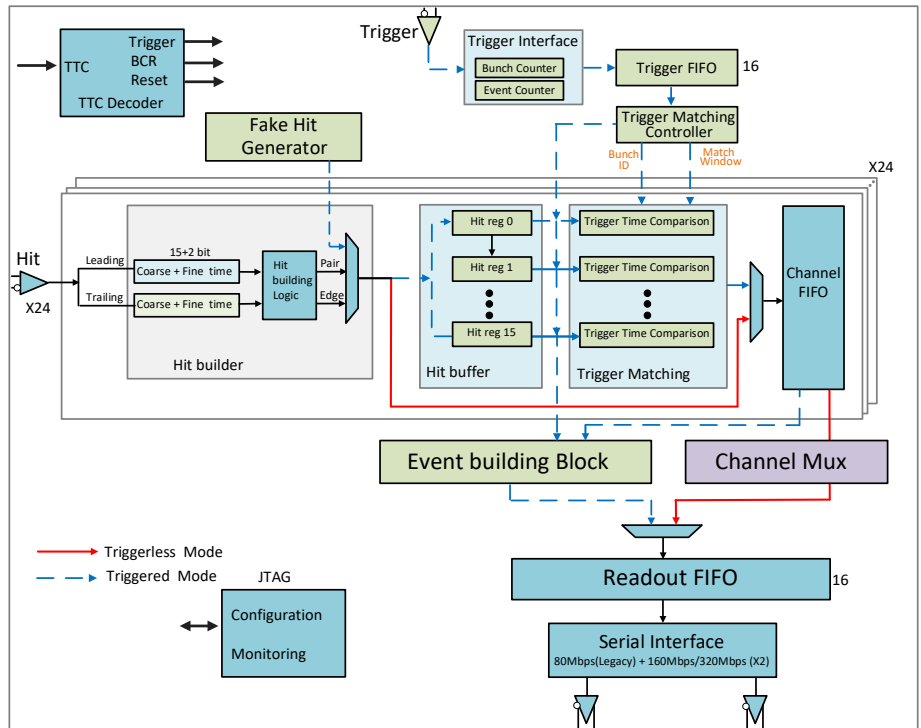

Fig. 4. Diagram for the TDC logic unit. Digitized time data from all 24 channels can be sent either in the triggerless mode (data flow follows the red lines) or in the triggered mode (data flow follows the dashed blue lines). Detailed descriptions can be found in the text.

*2.3.1 Triggerless operation of the TDC*

In triggerless operation, leading and trailing edge times are collected by the "Hit builder" block in each channel. Edge time measurements are assembled into separate packets in edge mode, while they are combined by attaching the time difference to the leading edge time in pair mode. Once a hit is built in the desired mode, it is transmitted to a 4-deep channel FIFO, from which the data from all 24 channels are multiplexed (Channel Mux) to a shared Readout FIFO of depth 16. Time measurements from the readout FIFO are further re-assembled in the "Serial Interface" for output. There are three output modes: 320/160 Mbps with two lines, and the legacy mode of 80 Mbps like in the AMT. In the 320/160 Mbps modes, the raw time measurements are encoded into three and four consecutive 8b/10b packets for edge and pair mode operation, respectively, and the even/odd bits are distributed to these two output lines accordingly. In case the readout FIFO is empty, commas are inserted. We also limit the maximum number of consecutive data packets without commas. This limit on packets without commas is set by programmable control registers. This limit is essential for back-end decoding circuits, as commas are utilized for tracking the boundary of encoded packets. In the legacy mode, the packets are sent only when the readout FIFO is not empty and each packet is encapsulated with start/stop bits. Otherwise the output line remains idle.

| Width | 5b | 2b | 17b | 8b | Total |
|---|---|---|---|---|---|
| Edge mode: Leading or trailing | Channel ID | Mode "00"/"01" | Edge-Time Measurement | | 24 bits |
| Pair mode: Leading + width | Channel ID | Mode "11" | Edge-Time Measurement | Pulse width | 32 bits |

Table 1. Data format for edge and pair mode of the TDC. For the edge mode, each edge is represented by 24 bits, leading to a total of 48 bits for a hit. In pair mode, 32 bits are needed to encode the combined time information of both edges. Pair mode is thus reducing the data volume to be transmitted to the CSM by about one third.

*2.3.2 Triggered operation of the TDC*

In triggered operation, the Hit builder in each channel is shared, and assembled hit packets in the selected mode (pair/edge mode) are buffered in a ring buffer with a depth of 16, awaiting trigger requests. A trigger request is introduced either from the TTC or through the dedicated Trigger input. The trigger is time-stamped with a local Bunch Crossing ID (BCID) counter, compensated by a time offset to represent the correct trigger time. The trigger time is then written into a trigger FIFO together with its event ID. The event ID is derived from a counter and incremented by one per trigger request.

Buffered triggers in the trigger FIFO are transmitted sequentially by a "Trigger Matching Controller" until the FIFO is empty. Explicit trigger matching information is generated in the form of trigger time (BCID) plus an expanded range for events matching (Matching Window). This information is distributed to each channel with a trigger request. Upon receipt of the trigger request, the ring buffers are loaded into the "Trigger Matching" unit in each channel. A time comparison is performed to find time measurements within the Matching Window, which are then sent to the Channel FIFO. The hits in the ring buffer are retained for a programmable time and during which they can be selected by multiple triggers. All hits belonging to a particular triggered event are arranged in the format of an event header, the matched hits, and a trailer. The event header is a 24-bit word including the trigger time and event ID, whereas the trailer is a 24-bit word comprising the number of matched hits and potential error flags, such as FIFO overflow. Matched hits in all channel FIFOs are read out sequentially to the event builder block and the number of hits is reported in the trailer. The serial output from the readout FIFO is identical to that of the triggerless operation. The information of header and trailer is encoded into three consecutive 8b/10b packets. Encoded matched hits between header and trailer follow the same format as the one of edge or pair mode in triggerless operation. Events from consecutive triggers are separated by commas.

A different trigger matching scheme is implemented in this prototype compared to the current AMT ASIC. In the AMT, all hits from 24 input channels are buffered in a L1 buffer with a depth of 256. The L1 buffer is implemented via a dual-port RAM, thus a state machine is required to keep track of the write and read addresses of the memory. A trigger matching algorithm is performed by operating the address pointers to search for hits within the time matching window.

There are several drawbacks of this scheme [14], for example, the L1 buffer is shared among all 24 channels, and thus it cannot guarantee that hits are written in the order of their time stamps, making it complicated to cover all events in the matching window correctly. Similarly, the address pointers need to be organized properly to flush old hits out that no longer can be selected for readout. In this TDC design, however, we deploy trigger matching for each individual channel. This reduces the total size of channel buffers and allows to use ring buffers instead of a dual-port RAM. Another advantage of this scheme is that hits are always written in time sequence and trigger matching can be performed by making comparison of all ring buffer contents simultaneously, instead of reading them out one by one via address pointers. Old hits are overwritten automatically by "Fake Hits", which are generated at a programmable rate. There are special tags in these fake events, preventing them from being considered as real hits during the trigger matching process.

## 3. Performance Characterization

### 3.1. TDC timing Performance

We characterize the timing performance of the TDC using uniformity of the fine-time bin size as well as its timing precision. The fine-time bin size is measured with the code density test [15], and ambiguous bins are corrected following the proposed scheme in [10]. There are two independent TDC slices in each channel, corresponding to the fine time measurements of the leading and trailing edges of a hit. This scheme was introduced to handle narrow pulses. A typical result for the four bins in all 48 TDC slices is shown in Figure 5, in which the horizontal axis is the TDC slice number (1-48), and the vertical axis represents the bin size in the corresponding slice. The mean bin size is found to be 781 ps with a variation of less than 40 ps (±5% of the bin size).

Figure 6 presents a scanning of the TDC timing performance as a function of the time interval. Tests were performed for all 48 TDC slices. The result agrees with the prediction of the time resolution induced solely by the quantization of the TDC fine-time bin size [16], indicating that the timing performance is close to the theoretical prediction and contributions from other sources are negligible. These results are comparable with those in the demonstrator prototype, implying that there are no performance degradations due to ePLL simplification, technology migration, or integration of the TDC logic unit.

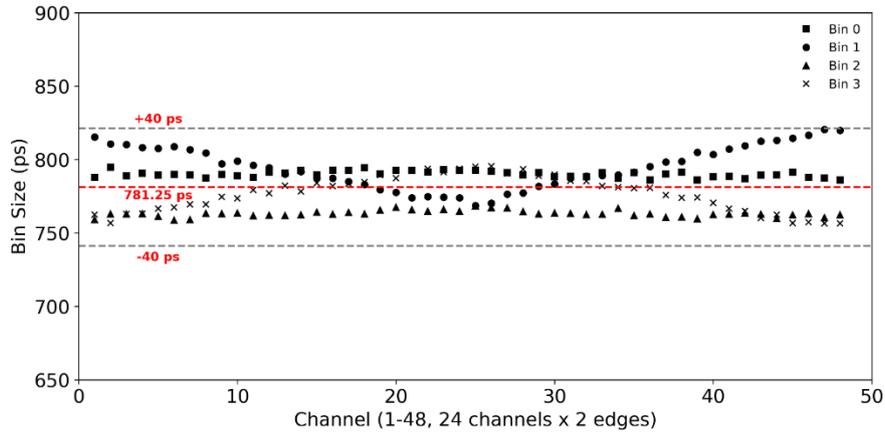

Fig. 5. Fine-time bin size for all 48 slices. The measurements are performed for each fine time bin separately.

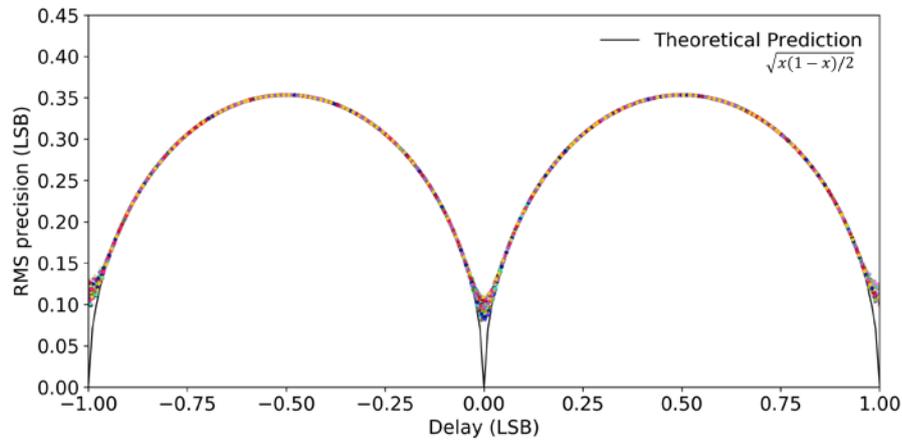

Fig. 6. The timing precision of 48 TDC slices with respect to different time intervals. The horizontal axis is the time interval measured by the TDC. The vertical axis is the RMS timing precision in units of the LSB.

*3.2. TDC logic unit performance*

To evaluate the performance of the TDC logic unit, the 24 channels of one TDC are connected to an FPGA evaluation board from which pre-defined patterns are generated for both triggerless and triggered operations. The TDC outputs are connected back to the FPGA, where the data are analyzed to verify chip functionality. In triggerless operation, the channel inputs are driven by pseudo-random hits with an average frequency of either 200 or 400 KHz. For random hit generation, the channel-to-channel dependence is minimized by adopting different seeds for the pseudo-random generator in each channel. The performance of the TDC logic unit in terms of processing time and reliability was verified by calculating the time delay through the TDC, the latency, for each event as well as the frequency of eventual packet loss. A distribution of the latency through the TDC for 320 Mbps outputs is shown in Figure 7, where Figure 7(a) presents the latency distribution and Figure 7(b) shows the percentage of received packets within a predefined latency cutoff time. The TDC was operating in pair mode, and no packets loss was observed during over 24 hours of continuous operations. Results show that > 99% of packets can be transmitted within a latency of 350 ns for a hit rate of 200 or 400 KHz per channel, respectively. With the maximum line rate of 320 Mbps, tests indicate that the TDC is able to sustain a hit rate of up to 660 KHz per channel without any packet loss.

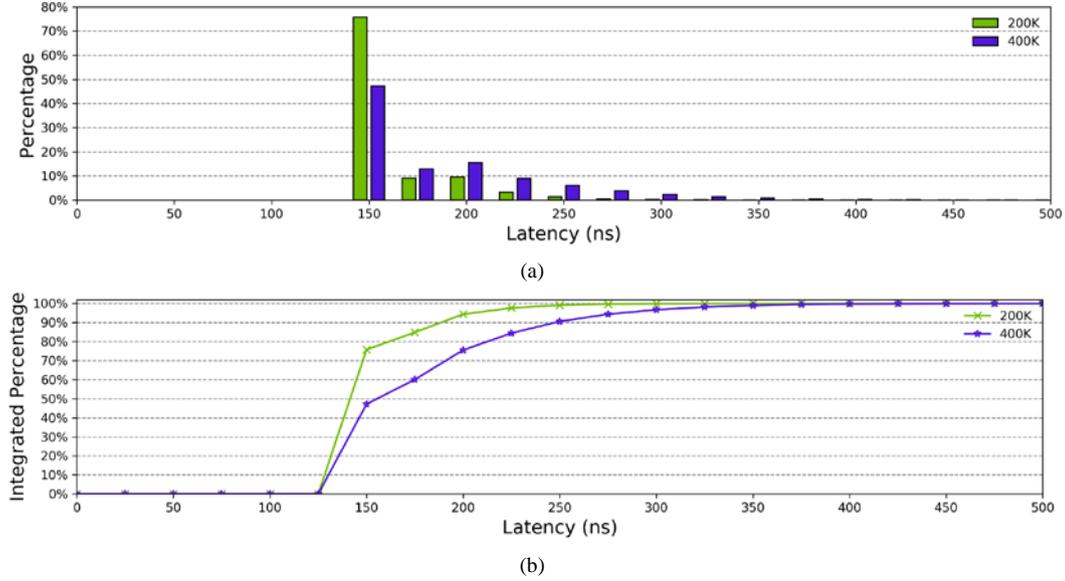

(a)

(b)

Fig. 7. (a) Latency distribution of the TDC in triggerless mode with an input rate of 200 and 400 KHz/channel; and (b) Percentage of data sent as a function of a given latency cut-off time. In these studies, the TDC works in pair mode with an output rate of 320 Mbps per line.

To test the triggered operation mode under "extreme" conditions, i.e. when instantaneous hit rates are far beyond expectation for normal LHC operation (as could be caused by EMI or strong external noise sources), the inputs to the TDC were loaded by "noise bursts", where many hits were sent to all 24 channels within a short time interval. In this scenario we verify the triggered operation by introducing trigger requests at different times with different matching window sizes with respect to the beginning of each burst group. The matched results are predictable for each trigger request, allowing for real-time packet checking inside the FPGA. No hang-ups due to buffer overflows or other malfunctions were observed.

*3.3. Power Consumption*

The TDC power consumption is found to be 250 mW. It is measured by connecting a 0.1 Ω resistor in series with the power supply and monitoring the current across the resistor. All 48 TDC slices are running during the measurement, and the total current is measured to be ~208 mA with a 1.2 volt supply. This includes about 86 mA from the digital and 122 mA from the analog parts of the chip. Compared to the demonstrator prototype, the reduction mainly has the following reasons: (1) the operating voltage is 1.2 volt in the TSMC process compared to the 1.5 volt in the GlobalFoundries process; (2) the simplified design of the ePLL leads to a reduction of the power consumption from about 30 mW to 12.5 mW according to circuit simulations; and (3) we use PTR instead of TSPC for the sampling registers used in the TDC fine time measurement unit, as described in section 2.2.

## 4. Summary

A prototype of the TDC ASIC containing all features for the upgrade of the readout electronics of the ATLAS MDT detector at the HL-LHC has been presented and fully tested. It is close to final as all implemented features were demonstrated and only triple modular redundancy for the controlling logic is pending before production. The operation modes of this ASIC include triggerless and triggered operation as well as configuration options for edge and pair mode and for output line rates. The ePLL and timing circuits have been optimized for power reduction. A detailed test program has been performed to verify timing behavior and latency as a function of the input data rate and, in case of the triggered mode, trigger rate and matching window size. The timing performance, measured over all 24 TDC channels, shows a uniformity of 5% of the fine time bin size inside each coarse time bin of 3.125 ns, largely exceeding the accuracy required for the MDT drift time measurement. The power consumption of the device is about 250 mW, corresponding to a power dissipation of about 10.4 mW per TDC channel, i.e. 5.2 mW per TDC slice. The measured latency of the TDC in the triggerless mode at an average hit rate of up to 400 KHz per channel demonstrates that > 99% of hits can be transmitted within 350 ns at an output line rate of 320 Mbps.

**Acknowledgments**

This work is supported by National Key Research Program of the Ministry of Science and Technology of China under grant 2016YFA0400100 and US National Science Foundation under contracts PHY01624739. We acknowledge the help from CERN micro-electronics group for providing the original ePLL design and S. Bugiel, *et al.* for providing the high-speed differential IOs.